\documentclass{article}

\usepackage{amssymb,graphicx}

\begin{document}

\title{\bf Dynamical Quantum Geometry \\ (DQG Programme)}
\author{Tim A. Koslowski\\{\texttt{tim@physik.uni-wuerzburg.de}}\\
Institut f\"ur Theoretische Physik und Astrophysik\\
Universit\"at W\"urzburg\\
Am Hubland\\
D-97074 W\"urzburg\\
European Union}
\date{\today}
\maketitle

\begin{abstract}
    In this brief note (written as a lengthy letter), we describe the construction of a representation for the Weyl-algebra underlying Loop Quantum Geometry constructed from a diffeomorphism variant state, which corresponds to a ''condensate'' of Loop Quantum Geometry, resembling a static spatial geometry. We present the kinematical GNS-representation and the gauge- and diffeomorphism invariant Hilbert space representation and show that the expectation values of the geometric operators take essentialy classical values plus quantum corrections, which is similar to a ''local condensate'' of quantum geometry. We describe the idea for the construction of a scale dependent asymptotic map into a family of scale dependent lattice gauge theories, where scale separates the essential geometry and a low energy effective theory, which is described as degrees of freedom in the lattice gauge theory. If this idea can be implemented then it is likely to turn out that this Hilbert space contains in addition to gravity also gauge coupled ''extra degrees of freedom'', which may not be dynamically irrelevant.
\end{abstract}

The algebra that underlies Loop Quantum Gravity is generated by matrix elements $h_e(A)_{IJ}$ of $SU(2)$-holonomies along arbitrary piecewise analytical curves $e$ in the Cauchy surface $\Sigma$ and the fluxes $E^i(S)$ of the conjugated electric fields through arbitrary piecewise analytical surfaces $S$. This algebra carries a canonical representation of the $SU(2)$-gauge transformations and the piecewise analytical diffeomorphisms on $\Sigma$. A $C^*$-algebra version of this algebra was introduced by Fleischhack \cite{fleischhack-weyl}. This algebra is constructed as a subalgebra of $\mathcal B(\mathcal H=L^2(\mathcal A,d\mu_{AL}))$, the bounded operators on the Hilbert space of w.r.t. the Ashtekar-Lewandowski measure square integrable functions on the quantum configuration space $\mathcal A$, generated by matrix elements of holonomies and Weyl operators corresponding to pull-backs under translations on $\mathcal A$, that are generated by the exponential action of the fluxes on $\mathcal A$. The gauge transformations and a subset of the homeomorphisms (for details see \cite{tim-diffeo}) are then implemented as pull-backs under homeomorphisms on $\mathcal A$, that preform the particular transformations on $\mathcal A$. The operators corresponding to the pull-backs are unitary due to the invariance of $d\mu_{AL}$ under Weyl-, gauge- and diffeomorphism-transformations.

It is the aim of this letter to construct a generalization of coherent states on the simple harmonic oscillator for a finite momentum $p_o$ in the sense of a finite spatial geometry $E_o$. These states will be outside the usual Hilbert space of Loop Quantum Gravity, and we will construct their GNS-Hilbert space and the GNS-representation of the algebra of Quantum Geometry thereon. The VEVs of the harmonic oscillator are $\omega_o(U_\lambda V_\mu) = \langle \Omega, U_\lambda V_\mu \Omega \rangle = e^{-(\frac 1 2\frac{\lambda}{ c})^2-i \lambda \mu -\frac{(c \mu)^2}{2}}$, where $U_\lambda=e^{i \lambda x}$ and $V_\mu=e^{i \mu p}$ denote the Weyl-operators. The expectation values in the coherent state $\Omega_\alpha=e^{\alpha a^*+\bar{\alpha} a}\Omega$, labeled by $\alpha=x_o+ip_o$ are for $x_o=0$:
$$
  \omega_\alpha(U_\lambda V_\mu) = e^{2ic p_o \mu} \omega_o(U_\lambda V_\mu).
$$
The generalization of this relation will turn out to be a state on the algebra of Quantum Geometry from which we construct a GNS-representation. This is the concern of the first part of this letter. After discussing some properties of this representation we conclude this letter with the introduction of a programme to construct an effective field theory for this representation, which is as of now an uncompleted programme, so we can only present the framework here.

It turns out that Fleischhacks Weyl-algebra is larger than what is needed for supporting Quantum Geometry: Quantum Geometry is provided by a family of operators for lengths, areas and volumes of paths, surfaces and regions in $\Sigma$. The length operator is constructed from the volume operator using Thiemann's trick \cite{thiemann-length}. The area operator for a surface $S$ is constructed \cite{al-area} from flux operators as
\begin{equation}
  A(S)=\lim_{\leftarrow \mathcal P(S)} \sum_{S_p \in\mathcal P(S)} \sqrt{\eta_{ij} E^i(S_p) E^j(S_p)},
\end{equation}
where $\mathcal P(S)$ runs over the partitions of $S$, which also have to include 1- and 0-dimensional quasi-surfaces for technical reasons. The resulting area operator turns out to be a sum over vertex-Laplace operators $A(S)\,Cyl_\alpha = \sum_{v \in S} 4 \pi \gamma l^2_{Pl}$ $\sum_v \sqrt{\Delta_{v,S,\alpha}}\,Cyl_\alpha$, where $\alpha$ is a graph that contains a vertex $v$ for all (transversal) intersections with $S$. The observation that makes the restriction to a subalgebra possible is that one can find a quantization of the volume of a region constructed from area operators, which is different from \cite{al-volume,rs-volume}, where an expression of the form $\int_R \sqrt{det(E)}$ is quantized. This suggests to consider the area operator as fundamental and some of the fluxes as composite operators.

The volume of a classical region is (if the metric is suitably regular) the limit of Riemann sums of the volume of cells of coordinate volume that approaches zero. For suitably regular regions $R$, we can find a sequence of coordinate-cubical complexes that approach the interior of $R$, such that the coordinate volume of the cells approaches zero. The metric inside each cell can be assumed to be homogeneous, e.g. a ''volume average'' over the cell. The volume of a cube in a homogeneous metric is however precisely the volume of a parallelepiped in Euclidean metric, after changing into the (in this case global) Riemann normal coordinate system. So, given a three-dimensional generalization of Herons formula of the area of a parallelogram in terms of the three independent length of the parallelogram to the three-dimensional case using the six independent areas on a parallelepiped to express its volume. The construction of this expression goes as follows: Using a rotation to a adapted coordinate system, the parallelepiped is spanned by $\vec a=(a_1,0,0)$, $\vec b=(b_1,b_2,0)$ and $\vec c=(c_1,c_2,c_3)$, so the volume is $V=a_1\,b_2\,c_3$. The system of equations involving the squares of the three surface areas $A_i$ and the squares of the areas of three diagonal cross sections $B_j$ is:
\begin{equation}\label{surface-equ}\small{
  \begin{array}{rcccl}
    {A_a} &=&|\vec b \times \vec c|^2&=& {\left( {b_2}{c_1} - {b_1}{c_2} \right) }^2 + \left( b_1^2 + b_2^2 \right)c_3^2\\
    {A_b} &=&|\vec a \times \vec c|^2&=& a_1^2\left( c_2^2 + c_3^2 \right)\\
    {A_c} &=&|\vec a \times \vec b|^2&=& a_1^2 b_2^2\\
    {B_a} &=&|(\vec b+\vec c)\times \vec a|^2&=& a_1^2\left( {\left( {b_2} + {c_2} \right) }^2 + c_3^2 \right) \\
    {B_b} &=&|(\vec a+\vec c)\times \vec b|^2&=& \left( {b_2}\left( {a_1} + {c_1} \right)  - {b_1}{c_2} \right)^2 +
    \left( b_1^2 + b_2^2 \right) c_3^2\\
    {B_c} &=&|(\vec a+\vec b)\times \vec c|^2&=& \left( {b_2}{c_1} - \left( {a_1} + {b_1} \right) {c_2} \right)^2 +
    \left( \left( {a_1} + {b_1} \right)^2 + b_2^2 \right) c_3^2.
  \end{array}}
\end{equation}
This coupled system of seven multi-linear equations can be solved for the volume $V$, which was aided by computer algebra:
\begin{equation}\label{explicit-volume}
  \small{
  \begin{array}{rcl}
    V_o &=& \biggl| \frac{a\,b^2\,c^2\,d\,{A_b}}
          {4\,\left( b^4\,d^2\,{{A_b}}^2 - 8\,b^2\,c^2\,d\,{A_b}\,{A_c} + 16\,c^4\,{{A_c}}^2 \right) } +
         \frac{b^4\,d\,e^2\,{A_b}}{16\,\left( b^4\,d^2\,{{A_b}}^2 - 8\,b^2\,c^2\,d\,{A_b}\,{A_c} + 16\,c^4\,{{A_c}}^2 \right) } \biggr.
         \\ &&-
         \frac{b^4\,d^2\,f\,{A_b}}{16\,\left( b^4\,d^2\,{{A_b}}^2 - 8\,b^2\,c^2\,d\,{A_b}\,{A_c} + 16\,c^4\,{{A_c}}^2 \right) } -
         \frac{a\,c^4\,{A_c}}{b^4\,d^2\,{{A_b}}^2 - 8\,b^2\,c^2\,d\,{A_b}\,{A_c} + 16\,c^4\,{{A_c}}^2}
         \\ &&+
         \frac{b^2\,c^2\,e^2\,{A_c}}{4\,\left( b^4\,d^2\,{{A_b}}^2 - 8\,b^2\,c^2\,d\,{A_b}\,{A_c} + 16\,c^4\,{{A_c}}^2 \right) } +
         \frac{b^2\,c^2\,d\,f\,{A_c}}{4\,\left( b^4\,d^2\,{{A_b}}^2 - 8\,b^2\,c^2\,d\,{A_b}\,{A_c} + 16\,c^4\,{{A_c}}^2 \right) }
         \\ &&\biggl. +
         \frac{{\sqrt{a\,b^6\,c^2\,d^2\,e^2\,{{A_b}}^2 - 4\,a\,b^4\,c^4\,d\,e^2\,{A_b}\,{A_c} + b^6\,c^2\,d\,e^4\,{A_b}\,{A_c} -
               b^6\,c^2\,d^2\,e^2\,f\,{A_b}\,{A_c} + 4\,b^4\,c^4\,d\,e^2\,f\,{{A_c}}^2}}}{4\,
            \left( b^4\,d^2\,{{A_b}}^2 - 8\,b^2\,c^2\,d\,{A_b}\,{A_c} + 16\,c^4\,{{A_c}}^2 \right) } \biggr| ^{\frac{1}{4}}
  \end{array}}
\end{equation}
where:
$$
  \small{\begin{array}{rcl}
     a &=& {\left( -{{A_b}}^2 - {\left( {A_c} - {B_a} \right) }^2 + 2\,{A_b}\,\left( {A_c} + {B_a} \right)  \right) \,
      \left( {{A_a}}^2 + {\left( {A_c} - {B_b} \right) }^2 - 2\,{A_a}\,\left( {A_c} + {B_b} \right)  \right) }\\
  b &=& {{\sqrt{{{A_b}}^2 + {\left( {A_c} - {B_a} \right) }^2 - 2\,{A_b}\,\left( {A_c} + {B_a} \right) }}}\\
  c &=& {{A_b}\,\left( {{A_b}}^2 + {\left( {A_c} - {B_a} \right) }^2 - 2\,{A_b}\,\left( {A_c} + {B_a} \right)  \right) }\\
  d &=& {{\left( {A_b} + {A_c} - {B_a} \right) }^2\,
      \left( {{A_b}}^2 + {\left( {A_c} - {B_a} \right) }^2 - 2\,{A_b}\,\left( {A_c} + {B_a} \right)  \right) }\\
  e &=& {{\left( -{{A_b}}^2 - {\left( {A_c} - {B_a} \right) }^2 + 2\,{A_b}\,\left( {A_c} + {B_a} \right)  \right) }^
       {\frac{3}{2}}\,\left( {A_a} + {A_b} - {B_c} \right) }\\
  f &=& {\left( -{{A_b}}^2 - {\left( {A_c} - {B_a} \right) }^2 + 2\,{A_b}\,\left( {A_c} + {B_a} \right)  \right) \,
      \left( {{A_a}}^2 + {\left( {A_b} - {B_c} \right) }^2 - 2\,{A_a}\,\left( {A_b} + {B_c} \right)  \right) }
  \end{array}}
$$
\begin{figure}\center
  \includegraphics[width=\textwidth]{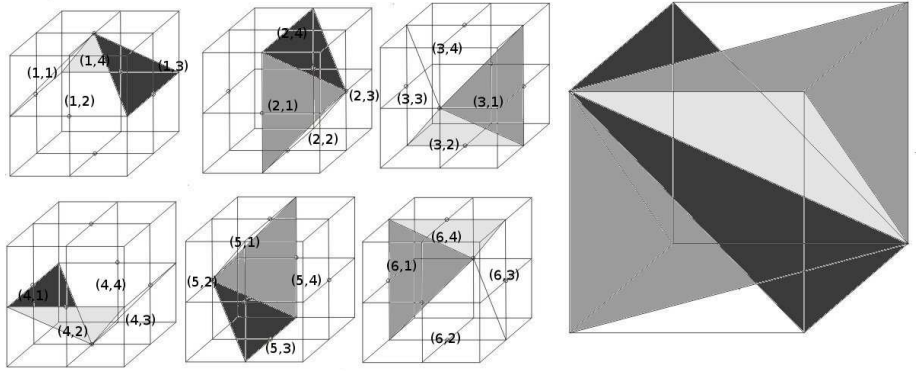}\\
  \caption{This figure illustrates the relation between the ''used diagonal cross sections'' and the ''used faces'' in the moved diagonals. The numbers indicate the pair $(i,j)$ in the labeling of the face. The coordinate chart $U$ is assumed to be right-handed cartesian and the 1-direction is assumed to be going from left to right.}\label{external-fig-2}
\end{figure}
We will modify the formula for $V$ by multiplying it with a factor $\theta(A_a,...,B_c)$ defined 1 if there is a non-degenerate parallelepiped with respective areas $A_a,...,B_c$ and 0 otherwise. The classically equivalent formula for the volume is then:
\begin{equation}\label{vol-func}
  V(A_a,...,B_c) := V_o(A_a,...,B_c) \theta(A_a,...,B_c).
\end{equation}
It turns out to be useful to not use the diagonals themselves, due to their intersection at the ''coordinate center of mass'', but to use isometric surfaces as indicated in figure \ref{external-fig-2}, which we call ''moved diogonals''. The classical volume does not depend on the particular choice of sequence of cubical decompositions, the quantum operator is however sensitive to it. Hence we provide a particular choice and then remove the finger prints of this choice by averaging over a suitable set of classically equivalent choices. Let us for simplicity assume that $R$ is contained in a single chart $(U,\phi)$, if not we have to use a partition of unity to achieve this. Then for all $\epsilon >0$ we can decompose $R$ into coordinate cubes $C^{U,\epsilon}_{\vec n}=\{\phi^{-1}(\vec x):n_i \epsilon \le x_i < (n_i+1) \epsilon\}$. Those cubes that are subsets of $R$ form a cubical decomposition $L_{U,\epsilon}(R)$. Clearly $V(R)=\lim_{\epsilon \rightarrow 0} \sum_{C\in L_{U,\epsilon}} V(A_a(C),...,B_c(C))$ converges to the classical volume of $R$ for all charts $U$. Moreover, removing any family $S_\epsilon$ of subsets from $L_{U,\epsilon}$ for which the coordinate volume vanishes as $\epsilon \rightarrow 0$ does not change the classical value of $V(R)$. To treat the surfaces democratically, we will insert for $A_i=\frac 1 2(A(S^f_i)+A(S^b_i))$, where $S^f_i$ is the ''front'' surface in $i$ direction and $S^b_i$ the respective ''back'' surface.

The strategy for the construction of the quantum operator is as follows: We construct an essentially self-adjoint volume operator on the spin network functions using the limit of $V(R)=\lim_{\epsilon \rightarrow 0} \sum_{C\in L_{U,\epsilon}} V(A_a(C),...,B_c(C))$ with $V$ as in equation \ref{vol-func} together with a suitable average of $L_{U,\epsilon}$ then define its Hermitian extension to $\mathcal H$ as the desired volume operator. We will in particular adapt the definition to the underlying graph: Given a spin network function $\psi_\alpha$ depending on a graph $\alpha$, there is a chart $(U,\phi)$ and a value for $\epsilon$, s.t. each cell contains at most one vertex, which is at the ''coordinate center of mass'' of the respective cell $C^{U,\epsilon}_{\vec n}$, and all cells that do not contain a vertex contain at most one edge, which is furthermore required that its restriction to the cell is connected. Moreover, this chart can be chosen s.t. the edges of $\alpha$ penetrate the surfaces $A_a(C),...,B_c(C)$ through its interior. We call one of these charts $(U_\alpha,\phi_\alpha)$. Given such a chart, we can define a refinement process, by subdividing each cell into $3 \times 3 \times 3$ cells, such that each vertex containing cell is divided into a central cell containing the vertex at its ''coordinate center of mass'' and 26 adjacent cells. These adjacent cells may have to be subdivided further into coordinate parallelepipeds, such that each cell contains at most one edge. This subdivision process lets us take the limit of the Riemann sums as a limit over the subdivisions procedure.

It is a consequence of the $\theta$-factor that each cell, that does not contain a vertex, will be assigned zero volume, because a single connected edge can not penetrate the boundary of a cell in three independent points, hence the parallelepiped is degenerate. Since the volume functional is additive, we have to only sum over the value of $V$ for vertex containing cells. Each cell can be treated separately:

The investigation of the vertex containing cells shows that the value of $V$ depends on the choice of chart through the topological relations of the edges adjacent to the vertex with the surfaces $A_a,...,B_c$ (i.e. which surfaces are penetrated by which set of edges). To remove this dependence we first notice that due to the restriction to piecewise analytic edges there is a value of $\epsilon$ such that all edges are ''outgoing from the vertex $v$'', but at the same time for any value $\epsilon >0$ there is a stratified diffeomorhism $\varphi(\alpha,v)$ such that the adjacent edges can be mapped to penetrate any of the surfaces $A_a,...,B_c$, and this diffeomorphism affects only the cell and its next neighbors. Thus, for each topological relation between the vertex-adjacent edges and surfaces $A_a,...,B_c$ there is a chart $U_\alpha^{\varphi(\alpha,v)}$ s.t. this topological relation is satisfied in the corresponding decomposition. Taking one representative out of each of these classes and averaging the volume functional over all possible topological relations removes the chart dependence and we arrive at a background independent volume functional. We denote these topological relations by $\mathcal T(v)$.

The final volume operator is defined as the extension by density of the operator acting on spin network functions $\psi_\alpha$:
\begin{equation}
  V(R) \psi_\alpha:=\sum_{v \in V(\alpha \cap R)} \frac{1}{|\mathcal T(v)|}\sum_{t \in \mathcal T(v)} V(A^2(S^t_{A_a}),...,A^2(S^t_{B_c}))\,\psi_\alpha,
\end{equation}
where $S^t_{A_a},...,S^t_{B_c}$ denotes the resp. surfaces that satisfy the topological relation $t$ and $A$ denotes the area operator. This operator is well defined because the spin network functions are simultaneous eigenfunctions of the area operators in our ''external'' regularization and is gauge invariant by inspection.

We are now able to define the restricted Weyl-algebra and we start with a more general consideration: Generalizing the notion of \cite{fleischhack-weyl}, we consider a compact Hausdorff space $\mathbb X$ and a regular Borel probability measure $\mu$ thereon. The integral kernels $K\,f:x \mapsto \int d\mu(x^\prime)K(x,x^\prime) f(x^\prime)$ that leave $C(\mathbb X)$ invariant and are invertible contain a group of ''unitary'' elements, i.e. those elements for which $\int d\mu(x) \overline{(K f_1)(x)}(K f_2)(x)=\int d\mu(x) \overline{ f_1(x)} f_2(x)$, denoted by $\mathcal U(\mu)$. The commutative $C^*$-algebra $C(\mathbb X)$ acts as multiplication operators on $L^2(\mathbb X,d\mu)$, whereas the elements of $\mathcal W$ act as convolutions. This canonical action is denoted by $\pi_o$. Fixing a subgroup $\mathcal W$ of $\mathcal U(\mu)$, one can define the *-algebra $\mathfrak A_o(\mathbb X,\mathcal W)$ of finite sums of ordered pairs:
$$
  \mathfrak A_o(\mathbb X,\mathcal W) \ni a = \sum_{i=1}^n f_i \circ w_i,
$$
where $f_i \in C(\mathbb X)$ and $w_i \in \mathcal W$. The involution is given by $a^*=\sum_{i=1}^n w_i^{-1}(f_i) w_i^*$, where $w(f): x \mapsto \int d\mu(x^\prime) w(x,x^\prime) f(x\prime)$. The canonical representation of this *-algebra on $L^2(\mathbb X,d\mu)$ can be completed to a $C^*$-algebra, denoted by $\mathfrak A(\mathbb X,\mathcal W)$, in the operator norm and $||f \circ 1|| \le ||f||_\infty$ and $||1 \circ w||=1$.

The construction of unitaries follows from a slight modification of continuous measure generating systems in \cite{fleischhack-weyl}: A {\bf Hermitian measure generating system} is a set $E$ of continuous functions that is dense in $C(\mathbb X)$ as well as $L^2(\mathbb X,d\mu)$ containing $1$, such that all elements of $E\setminus \{1\}$ are perpendicular to $1$ in the inner product of $L^2(\mathbb X,d\mu)$, together with a labeling set $H$, consisting of a commuting set Hermitian operators in $\mathcal B(L^2(\mathbb X,d\mu))$ that have the elements of $E$ as eigenfunctions and their eigenvalues distinguish the elements of $E$. Without loss of generality, we assume that $h\,1=0 \forall h \in H$. For any analytic function $f$ on the eigenvalues of the labeling set$H(E)$ we define the linear map
$$
  \tau_f e := e \exp(i f(H(e))),
$$
as the extension by density to $L^2(\mathbb X,\mu)$, where $H(e)$ denotes the set of eigenvalues of $e\in E$. It turns out that if $f(H(1))=0$ then $\tau_f$ defines a unitary element of $\mathcal B(L^2(\mathbb X,d\mu))$.

Since $\mathfrak A_o(\mathbb X,\mathcal W)$ is dense in $\mathfrak A(\mathbb X,\mathcal W)$, we define a state on $\mathfrak A_o$, check whether it is bounded and extend it by density to $\mathfrak A$: Given a continuous morphism from $\mathcal W$ to $U(1)$, we define a functional $\omega_{F,\mu}$ on $\mathfrak A_o$ by:
\begin{equation}
  \omega_{F,\mu}\left(\sum_{i=1}^n f_i w_i\right):= \sum_{i=1}^n F(w_i) \int d\mu(x) f_i(x).
\end{equation}
This functional is bounded $|\omega_{F,\mu}(a)|\le \sum_{i=1}^n ||f_i||_\infty$ and linear as is checked by insertion. Positivity can be shown using the group morphism property of $F$ and the invariance of $\mu$ under the action of $\mathcal W$, such that
$$
  \begin{array}{rcl}
    \omega_{F,\nu}(a^* a) &=& \int d\mu(x) \sum_{i,j=1}^n \overline{(F(w_i) f_i)(x)} (F(w_j) f_j)(x)\\
                          &=& \int d\mu(x) \left|\sum_{i=1}^n f_i(x)\right|^2 \ge 0.
  \end{array}
$$
To calculate the GNS-representation constructed from $\omega_{F,\mu}$ it is useful to relate this to the canonical representation, usually constructed from the state $\omega_o(a):=\sum_{i=1}^n \int d\mu(x) f_i(x)$, using the map
\begin{equation}
  \kappa_F: a \mapsto \sum_{i=1}^n (F(w_i)^* f_i)\circ w_i.
\end{equation}
It turns out that $\kappa_F$ is an automorphism of $\mathfrak A_o$ and that $\kappa_f(a)$ is in the Gel'fand ideal $\mathcal I_F$ of $\omega_{F,\mu}$, if and only if $a$ is in the Gel'fand ideal $\mathcal I_o$ of $\omega_o$ and this relation can be carried over to $\mathfrak A$ due to the boundedness of $\kappa_F$. It follows that a dense set of representatives $\mathcal R_F$ of $\mathcal N_F$ in $\mathfrak A$, where $\mathcal N_F:=\mathfrak A/\mathcal I_F$, implies a set of representatives $\mathcal R_o$ and vice versa. This allows us to use a continuous generating system for the canonical representation as a dense set in the GNS-Hilbert space constructed from $\omega_{F,\mu}$. We notice that a weakly continuously represented 1-parameter group in the canonical representation implies an weakly continuous representation of this 1-parameter group in the GNS-representation due to the group morphism property of $F$ and due to the assumed continuity of $F$.

Fleischhacks definition of the Weyl-algebra of Loop Quantum Gravity fits precisely into this setup: The compact Hausdorff space is space of generalized connections $\mathcal A=\lim_{\alpha} (\mathcal P_\alpha, SU(2))$, where $\mathcal P_\alpha$ is the path-groupoid of the graph $\alpha$ and the projective limit is taken over the inclusion as subgraphs. The translations $W^i(S)$ on $\mathcal A$ generated by the fluxes $E^i(S)$ are special invertible convolution operators, that leave the Ashtekar-Lewandowski measure invariant. The definition \cite{fleischhack-weyl} of $\mathfrak A$ then coincides with our definition when taking $\mathbb X=\mathcal A$, $\mu=\mu_{AL}$ and $\mathcal W$ is the Weyl group generated by the $W^i(S)$ through arbitrary quasi-surfaces. The canonical state $\omega_o(Cyl\circ W)= \int d\mu_{AL} Cyl(A)$, where $Cyl \in Cyl(\mathcal A)$ and $W$ is an element of the Weyl group. The extended analytic diffeomorphisms leave the Ashtekar-Lewandowski measure invariant and hence the pull-backs under these diffeomorphisms define unitary operators in $\mathcal B(\mathcal H)$.

The gauge-variant spin network functions are a continuous measure generating system. A Hermitian labeling set for this generating system is constructed as follows: Given a map $\tau: \Sigma \rightarrow SU(2)$, we can consider the fluxes ''parallel to $\tau$'', i.e. $E_{||\tau}(S) := \int_S E^i\tau_i$. It turns out that the ''fluxes parallel to $\tau$'' together with the area operators form a Hermitian labeling set for the gauge variant spin network functions.

Let $\tau: \Sigma \rightarrow SU(2)$ be normalized, i.e. $k_{ij}\tau^i\tau^j=1$ at each point. Let us define three families of Weyl-operators for each oriented surface through their action on gauge variant spin network functions $\psi_\alpha$:
\begin{equation}
  \begin{array}{rcccl}
    W^A_S(f) &:&T&\mapsto& \alpha^A_{S,\lambda(f)} (T)\\
    W^+_S(f) &:&T&\mapsto& (\Theta_{S,\sigma,\exp(\lambda(f)\tau)})^* T\\
    W^-_S(f) &:&T&\mapsto& (\Theta_{S,\overline{\sigma},\exp(\lambda(f)\tau)})^* T,
  \end{array}
\end{equation}
where $  W^A_S(\lambda) \psi_\alpha := e^{i \lambda A_S} \psi_\alpha = \sum_i a_i T_{\gamma_S,i} e^{i \lambda A_S(\psi_{\alpha S,i})}=:\alpha^A_{S,\lambda}(\psi_\alpha)$ and $\Theta$ denotes the respective element of Fleischacks translation group on $\mathcal A$ corresponding to the exponential action of the respective flux parallel to $\tau$ with the orientation function $\sigma$ on the quasi-surface $S$ resp. the flipped orientation function $\overline{\sigma}$. The extension to operators in $\mathcal B(\mathcal H)$ by density defines unitary operators, which are precisely of the kind described above, because their integral kernel can be obtained through exponentiating a function on the labeling set for which $f(1)=0$. The group $\mathcal W_\tau$ generated by these operators contains area- and flux-Weyl-operators on less than two-dimensional quasi-surfaces, but they all flux-Weyl-operators are parallel to $\tau$. The algebra that we consider is then the algebra in $\mathcal B(\mathcal H)$ generated by the finite sums
$$
  \mathfrak A_\tau \ni a \sum_{i=1}^n Cyl_i W_i,
$$
where $Cyl_i in Cyl(\mathcal A)$ and $W_i \in \mathcal W_\tau$.
Let us fix a regular densitized inverse triad $E_o$. The state $\omega_{E_o}$ on the finite sums is:
\begin{equation}\label{Eo-state}
  \omega_{E_o}(a):= \sum_{i=1}^n F_{E_o}(w_i)\int_{\mathcal A} d\mu_o(A) f(A),
\end{equation}
where $F_{E_o}(W_S)=1$ whenever $S$ is contained in a less than two-dimensional subset of $\Sigma$ and otherwise
\begin{equation}\label{F-defi}
  F_{E_o} : \left\{
    \begin{array}{rcl}
      W^A_S(\lambda) & \mapsto & \exp(i \lambda \int_S |E_o|)\\
      W^+_S(\lambda) & \mapsto & \exp(i \lambda \int_S E_o)\\
      W^-_S(\lambda) & \mapsto & \exp(i \lambda \int_{\overline{S}} E_o).
    \end{array}
  \right.
\end{equation}
Let us use the automorphism $\kappa_F$ to describe the GNS-representation in terms of the canonical representation, for which we have:
$$
  \begin{array}{rcl}
    \omega_o(a_1^* a_2 a_3) & = & \langle \eta_o(a_1), \pi_o(a_2) \eta_o(a_3) \rangle_o\\
    \eta_o(a) : & A \mapsto & \sum_i f_i \\
    \pi_o(a) \phi: & A \mapsto & \sum_i f_i(A) (\alpha_{w_i}(\phi))(A)\\
    \langle \phi, \phi^\prime \rangle_o &:=& \int_{\mathcal A} \overline{\phi}(A) \phi^\prime(A),
  \end{array}
$$
Since $\omega_{E_o}(\kappa_F(a))=\omega_o(a)$, we see obtain the GNS-representation for $\omega_{E_o}$ immediately:
\begin{equation}
  \begin{array}{rccl}
    \eta_{E_o}(a) &= \eta_o(\kappa_F(a)) : & A \mapsto & \sum_i F(w_i) f_i \\
    \pi_{E_o} (a) \phi &= \pi_o(\kappa_F(a)): & A \mapsto & \sum_i F(w_i) f_i(A) (\alpha_{w_i}(\phi))(A)\\
    \langle \phi, \phi^\prime \rangle_{E_o} &=\langle \phi, \phi^\prime \rangle_o&:=& \int_{\mathcal A} \overline{\phi}(A) \phi^\prime(A).
  \end{array}
\end{equation}
The canonical action of the diffeomorphisms on the algebra elements moves the graphs $\gamma_i$ and the quasi-surfaces $S_i$ around:
$$
  \alpha_\phi(a) = \alpha_\phi(\sum_i f_{i,\gamma_i} w_{i,S_i}) = \sum_i f_{i \phi(\gamma_i)} w_{i,\phi(S_i)}.
$$
This implies for the transformation of the vacuum vector $\Omega_{E_o}$ of the GNS-representation:
$$
  \begin{array}{rcl}
    \omega_{E_o}(a) &=& \langle U_\phi \Omega_{E_o}, U_\phi \pi_o(a) \Omega_{E_o} \rangle_{E_o} \\
                    &=& \langle U_\phi \Omega_{E_o}, \pi_o(\alpha_\phi(a)) U_\phi \Omega_{E_o} \rangle_{E_o} = \omega^\phi_{E_o} (\alpha_\phi(a)),
  \end{array}
$$
The state $\omega^\phi_{E_o}$ is then determined to coincide with $a \mapsto \langle \Omega_{\phi(E_o)}, \pi_{\phi(E_o)}(a) \Omega_{\phi(E_o)}\rangle$, and we thus have the relation for the vacuum vectors:
\begin{equation}
  U_\phi \Omega_{E_o} = \Omega_{\phi(E_o)}.
\end{equation}
To be able to define a unitary action of the diffeomorphisms, we need to take the direct sums over all $E_o$ in the diffeomorphism orbit of $E_o$. Using the precise analogue of this calculation, we have for gauge transformations, which act on triads as $\alpha_\Lambda E =\Lambda^{-1} E \Lambda$, the relation
$$
  U_\Lambda \Omega_{E_o} = \Omega_{\alpha_\Lambda(E_o)}.
$$
Thus, to have a Hilbert space representation that carries a unitray representation of the diffeomorphisms and the gauge transformations, we should take the direct sum over the entire geometric orbit of $E_o$. There is one caveat concerning the map $\tau$: To be able to use group averaging, we have to have the fitting gauge transformed flux-Weyl-operators available in each summand, thus we need to take the direct sum over the GNS-representations with gauge transformed $\tau$-maps. The kinematical Hilbert space representation is thus the sum over the entire geometrical orbit $\mathcal G(E_o)$ of $E_o$:
\begin{equation}
    (\mathcal H_{\mathcal G_o},\pi_{\mathcal G_o,\tau}):= \oplus_{\{E_o:\mathcal G(E_o)=\mathcal G_o\}} (\mathcal H_{E_o},\pi_{E_o,\tau^G}),
\end{equation}
where $\tau^G$ denotes the gauge transformed $\tau$-map. The gauge-variant spin network functions are orthogonal in each of the $E_o$-representations, because for any two gauge-variant spin network functions $T_1,T_2$ we have:
$$
  \begin{array}{rcl}
  \langle T_1,T_2\rangle_{E_o}&=&\langle \pi(T_1) \Omega_{E_o},\pi(T_2) \Omega_{E_o} \rangle_{E_o} \\
    &=& \omega_{E_o}(T_1^* T_2)=\int_{\mathcal A} d\mu_o(A) \overline{T_1(A)} T_2(A) \\
    &=& \omega_o(T_1^* T_2) = \langle T_1, T_2 \rangle_o
  \end{array}
$$
This implies the kinematical orthogonality for normalized spin network functions $T_1,T_2$:
\begin{equation}
  \langle T_1 \circ E^1_o , T_2 \circ E^2_o \rangle_{\mathcal G_o} =\left\{
    \begin{array}{ll}
      1 &\textrm{for: }\, T_1=T_2 \,\wedge\, E^1_o=E^2_o \\
      0 &\textrm{otherwise.}
    \end{array}\right. ,
\end{equation}
so a complete orthogonal set is labeled by a gauge variant spin network function and a background geometry $E$ in the geometrical orbit of $E_o$.

The orthogonality of the different summands in the direct sum allows us to split the group averaging into three parts: First we classify the gauge invariant couplings between the spin network function and $E$, then we average over the group of transformations that affect the coupled spin network non-trivially and finally we average over the quotient of the gauge-transformation group by the group that acts non-trivially on the gauge-invariant spin network function, which will evidently leave the invariant couplings between the spin network and the background invariant. The treatment of the diffeomorphism constraint will be analogous:

Let us start with the gauge-invariant couplings: The basic observation that we need is that for edges $(y,x),(x,z)$ with $f(y,x)=x=i(x,z)$ objects of the form:
$$
  O^a:= Tr\left(((someth.)(A,E_o))(y,z) (h_{(y,x)}(A)) {E_o}^a_i(x) \tau^i (h_{x,z}(A))\right)
$$
are gauge invariant, because $(h_e(A))_{n,m} \mapsto \left(\Lambda^{-1}(i(e)) h_e(A) \Lambda(f(e))\right)_{n,m}$ and $E_o(x) \mapsto \Lambda^{-1}(x) E_o(x) \Lambda(x)$, but obviously not diffeomorphism invariant. This is however only a special case of the general picture: Any function $F_x(E_o)$ built from $E_o(x)$ that transforms under some representation of $SU(2)$ can be gauge invariantly coupled to a spin network function $T$ with vertex $x$ by constructing a gauge-invariant intertwiner between the representation of $F_x(E_o)$ and representations adjacent to $x$ in $T$. Thus, given a gauge-variant spin network $T_\gamma$ on a graph $\gamma$, then we can couple it gauge invariantly to $E_o$ by assigning a function $F_v(E_o)$ and an gauge-invariant intertwiner $M_v$ between the representation of $F_v(E_o)$ and adjacent spins to each vertex $v$ of $\gamma$. We call spin network functions with invariant couplings to the background gauge-invariantly coupled spin network functions. 

Let us now average over all gauge transformations that act nontrivially on the gauge-invariantly coupled spin network function $\pi(T_\alpha(A,E_o))\Omega_{E_o}$: The solution (to this group averaging over the finite number of copies of $SU(2)$, one for each non-invariantly coupled vertex) is as in Loop Quantum Gravity given by the product states of traces over closed loops, with the addition that there may be vertices in these closed loops, which represent gauge-invariant couplings. So, basically the solution space to the Gauss-constraint is enlarged by spin-transfer between the spin network function and the geometric background. We call these solutions {\bf gauge-invariantly coupled gauge-invariant} spin network functions. Given a gauge-variant cylindrical function $Cyl$, we denote its group average by $G(Cyl):=\int \prod_{v \in var(\alpha)} d\mu_H(g_v) Cyl(...,g_{v_1}^{-1} h_{v_1,v_2} g_{v_2},...)$. Notice that the gauge-orbit of two distinct gauge-invariant couplings can yield the same group average, when the ''transferred spin'' of the two couplings equal.

Finally, we preform the group averaging over the quotient of the gauge transformations by the finite group that acts nontrivially on the gauge-invariantly coupled spin network functions, which are precisely those gauge transformations $\Lambda(v)=1_{SU(2)}$ for all non-invariantly coupled vertices $v\in var(\alpha)$: Given a gauge-invariantly coupled gauge-invariant spin network function $\pi(T_\alpha(A,E_o))\Omega_{E_o}$, these transformations act $\Lambda: \pi(T_\alpha(A,E_o))\Omega_{E_o}\mapsto \pi(T_\alpha(A,E_o))\Omega_{\Lambda^{-1}E_o\Lambda}$. With these preparations we can calculate the effect of the gauge-rigging map
$$
  \eta(\pi(T(E_o))\Omega_{E_o}): \pi(T^\prime(E^\prime_o))\Omega_{E^\prime_o} \mapsto \sum_{\Lambda \in n\mathcal G_{T,E_o}} \langle U_{\Lambda} \pi(T(E_o))\Omega_{E_o}, \pi(T^\prime(E^\prime_o))\Omega_{E^\prime_o} \rangle_{\mathcal G_o},
$$
and upon preforming the aforementioned three-step we obtain the closed expression for the gauge-invariant inner product:
\begin{equation}
  \begin{array}{l}
    \langle \eta(\pi(T(E_o))\Omega_{E_o}), \eta(\pi(T^\prime(E^\prime_o))\Omega_{E^\prime_o}) \rangle_{\mathcal O(E_o)}\\:=
      \eta(\pi(T(E_o))\Omega_{E_o})[\pi(T^\prime(E^\prime_o))\Omega_{E^\prime_o}] \\
      = \left\{
        \begin{array}{cl}
          \int_{\mathcal A} d\mu_o(A) \overline{G(T(E_o,A))} G(T^\prime(E_o,A)) & \textrm{for} E_o^\prime \in \mathcal O(E_o)\\
          0 & \textrm{otherwise}.
        \end{array}
      \right.
  \end{array}
\end{equation}
The gauge-invariant Hilbert-space is thus spanned by gauge-invariantly coupled gauge-invariant spin network functions, which are embedded into a gauge-orbit of a background $E_o$. 

The precise same line of reasoning can be applied to solving the diffeomorphism constraint: The diffeomorphism invariant couplings between the spin network-functions and the background are gauge- and diffeomorphism-invariant couplings between the pin network function and the background. The closed expression for the gauge- and diffeomorphism-invariant inner product turns out to be:
\begin{equation}
  \begin{array}{l}
    \langle \eta_{diff}(\pi(T_\gamma)\Omega_{\mathcal O(E_o)}), \eta_{diff}(\pi(T^\prime_{\gamma^\prime})\Omega_{\mathcal O(E^\prime_o)}) \rangle_{\mathcal G_o}\\
    =\left\{
      \begin{array}{cl}
        \sum_{\phi \in Sym(\mathcal O(E_o))}\int_{\mathcal A} d\mu_o(A) \overline{T_{\phi(\gamma)}(A)} T^\prime_{\gamma^\prime}(A) & \textrm{for: }\, \mathcal O(E^\prime_o) \in \mathcal G(\mathcal O(E_o))\\
        0 & \textrm{otherwise},
      \end{array}
    \right.
  \end{array}
\end{equation}
where $Sym(\mathcal O(E_o))$ denotes the subgroup of the diffeomorphisms that contains the symmetries of $\mathcal O(E_o)$. So the gauge-and diffeomorphism invariant Hilbert space consists of gauge-and diffeomorphism-invariantly coupled gauge-invariant spin networks, which are embedded into a geometry (modulus isometries of this geometry).
\begin{figure}\center
  \includegraphics[width=3cm]{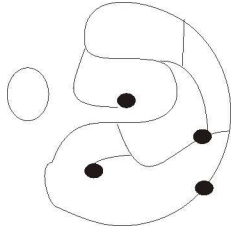}\\
  \caption{An example of a solution to the Gauss- and difeomrophism constraint: The graph is embedded up to isometries of the background, contains additional couplings to the background (black dots) and disconnected regions of the graph are not physically disconnected, due to the occurrence of the background geometry.}
\end{figure}

Having a Hilbert-space, that is spanned by spin-networks that are embedded into a background geometry $G_o$, one can consider the spin network as quantum fluctuations around this geometry. Let us make this statement more precise: 

Given a state $\omega$ on the algebra $\mathfrak A_\tau$ of Quantum Geometry, we call a countable set of zero- and one-dimensional embedded piecewise analytic submanifolds of $\Sigma$ the excess $\mathfrak E(\omega)$, if the expectation values of any area- or volume-operator on $\Sigma\setminus \mathfrak E(\omega)$ do not change upon removing a finite number of one-dimensional submanifolds form $\Sigma$. Since one can reconstruct a classical geometry form the areas and volumes of the embedded two- and three-dimensional submanifolds and since this geometry is invariant under the removal of a countable number of lower dimensional submanifolds, we are able to to define the the {\bf essential geometry} of a state $\omega$ as the geometry reconstructed from the expectation values of the area $\omega(A(S\setminus \mathfrak E(\omega)))$ and volume operators $\omega(V(R\setminus \mathfrak E(\omega)))$ on $\Sigma$. The essential geometry turns out to be a feature of the entire representation, because any sequence of cylindrical functions will be defined a countable set of graphs and thus the removal of this set of graphs will remove the excess of any element of the GNS representation and the removed set consists of a countable number of zero- and one-dimensional submanifolds. It thus turns out that the essential geometry of a state coincides with the essential geometry of any other state in the same GNS representation and are thus an invariant of the representation. So one can calculate the essential geometry of any state from the essential geometry of the ground state, which does not have an effect (due to the regularity assumption on $E_o$) and hence the essential geometry is simply reconstructed from the vacuum expectation value of the respective geometric operators.

The essential expectation values for the area operators on a surface $S$ are easily calculated as 
$$\begin{array}{rcl}
  \langle A(S)\rangle_{E_o}&=&\lim_{t\rightarrow 0}\frac{1}{2t}\left(\omega_{E_o}(W_A^S(t))-\omega_{E_o}(W_A^S(-t))\right) \\
  &=&\frac \partial{\partial t}\exp(t A_{E_o(S)})|_{t=0}=A_{E_o}(S),
  \end{array}
$$ 
so the essential expectation values of the area operators coincide with the classical areas of $S$ calculated in the geometry described through $E_o$. Calculating higher derivatives reveals that there are no fluctuations in the essential expectation values for the area operators. Moreover it turns out that the action on the ground state $\Omega_{E_o}$ of any two all area operators commute. This allows us to calculate the essential expectatioon values of the volume operator without further effort: The expectation values for the volume operator of a region is
$$
  \begin{array}{rcl}
    \langle V(R) \rangle_{E_o} &=& V(R)=\langle \Omega_{E_o},\lim_{\epsilon \rightarrow 0} \sum_{C\in L_{U,\epsilon}} V(A_a(C),...,B_c(C)) \Omega_{E_o} \rangle\\
      &=& V_{E_o}(R),
  \end{array}
$$
which is independent of the choice of chart $(U,\phi)$. Thus, the essential geometry turns out to be precisely the geometry that is described through the classical densitized inverse triad $E_o$. 

Since the essential geometry can be recovered from any state in the GNS-representation and is fixed by the $E_o$-geometry, we have a geometric background in the $E_o$-geometry that can be determined operationally, since the effect of a state can be determined operationally: Consider the following family of pairs of surfaces $\{(S,S\setminus x): S \in \mathcal S(\Sigma), x \in S\}$, where $\mathcal S(\Sigma)$ denotes the set of piecewise analytical surfaces in $\Sigma$. Moreover, consider the set of pairs of regions $\{(R,R\setminus x):R \in \mathcal R(\Sigma), x \in R\}$, where $\mathcal R(\Sigma)$ denotes the set of piecewise analytical regions in $\Sigma$. If the area- resp. volume- expectation values of any of the pairs disagree, then $x$ is in the effect of the state, so the effect of the state can be determined operationally, which means that there is a measurable geometric background that we can use to define essential distances, essential length of curves, essential areas, essential volumes and so on.

Using the essential geometry one can in particular measure the length of the edges of a cubical decomposition of compact subsets $C \subset \Sigma$. Each cubical decomposition has a dual graph with (at most\footnote{The vertices inside cells adjacent to the boundary of the compact region may have less than valence six.}) six-valent vertices. It turns out that the finite cubical decompositions of $\Sigma$ form a category {\bf Cub} with refinements as morphisms. A refinement of a cubical decomposition $\mathcal D_o(C)$ is a cubical decomposition $\mathcal D_1(C)$ that contains a (possibly trivial) decomposition for each cell $c \in \mathcal D_o(C)$. This category is partially ordered by the refinement property. Since there exists a cubical decomposition $\mathcal D_3(C)$ for any pair $\mathcal D_1(C),\mathcal D_2(C)$ that is finer than these two, they are furthermore a projective family. We will furthermore make use of the essential geometry and assume that the cubical decomposition has contains only edge-length between $l_o$ and $2 l_o$, meaning $\mathcal D_i$ is ''close to regular''.

To each cubical decomposition $\mathcal D(C)$, there is an embedded dual graph $\Gamma(\mathcal D)$ with generally six-valent vertices, which is constructed as follows: the vertices of $\Gamma$ are given by the ''coordinate center of mass'' of the cells and the links of $\Gamma$ are given by the concatenation of the geodesics from the coordinate center of mass of a cell to the coordinate center of mass of a joint face. (This construction is possible due to the occurrence of the essential geometry.) One can define the refinement of a lattice $\Gamma(\mathcal D_o)$ to a finer lattice $\Gamma(\mathcal D_1)$ through the the existence of a pair $\mathcal D_o,\mathcal D_1$, such that the cubical decompositions $\mathcal D_1 \ge \mathcal D_o$ are dual to $\Gamma_1$ and $\Gamma_2$ respectively. This turns the set of graphs that are dual to cubical decompositions into a category itself with refinements as morphisms. 

Given any graph $\Gamma$, we can consider the lattice gauge theory $LGT(\Gamma)$. The basic idea to construct an effective quantum field theory rests on the common belief that a Quantum Gauge Field Theory is the limit of lattice size going to zero of a lattice gauge theory\footnote{Notice that the metric limit is taken here, which is not possible in Loop Quantum Gravity, where only a projective limit over all cubical decompositions would be meaningful.} if this limit exists. If this limit does not exist, then one can at least call the family of lattice gauge theories a family of effective field theories. Thus, if one succeeds with the construction of a functor that takes a cubical decomposition $\mathcal D(C)$ into a lattice gauge theory $LGT(\Gamma(\mathcal D(C)))$, that encodes ''all relevant degrees of freedom''\footnote{We will at first map all degrees of freedom into the effective field theory and then allow to ''forget irrelevant degrees of freedom''.}, then one has constructed an effective field theory. Before we give a candidate construction for such a functor, we need to consider the construction of a contravariant functor that assigns a noncommutative $C^*$-algebra (the quantum algebra of an $SU(2)$-lattice gauge theory plus possible extra degrees of freedom) and a Hilbert-space representation thereof (the canonical representation of this lattice gauge theory on $L^2(SU(2)^{|\Gamma|},d\mu_H)\oplus \mathcal H_{extra}$):

Let $\mathcal D_o\le \mathcal D_1$ be two cubical decompositions of a compact subset $C\subset \Sigma$. Thus, for each cell $c\in \mathcal D_o$ there is a set of cells $r(c) \in \mathcal D_1$ that constitute a decomposition of $c$. Moreover, for each face $f \in \mathcal D_o$ there is a set of faces $r(f) \in \mathcal D_1$ that constitute a decomposition of the face $f$. Given a lattice gauge theory on $\Gamma(\mathcal D_1)$, we map all degrees of freedom on the $n$ links across the faces $r(f)$ into the degrees of freedom on the link across $f$. Moreover, we map all degrees od freedom residing on the vertices at the centers of the cells in $r(c)$ into the degrees of freedom on the vertex at the center of $c$. Since there is no bound on the number of cells in the refinement $r(c)$ of a cell $c$ and no bound on the number of faces in the refinement $r(f)$ of a face $f$, there is an infinite number of degrees of freedom on each link and each vertex. We denote this map from the finer $LGT(\Gamma_f)$ to the coarser lattice gauge theory $LGT(\Gamma_c)$ by $R_{fc}$. The consistency condition for the map $R$ is:
\begin{equation}\label{consistency}
  R_{23}(R_{12}(LGT(\Gamma_1)))=R_{13}(LGT(\Gamma_1)),
\end{equation}
whenever $\Gamma_1\ge\Gamma_2\ge\Gamma_3$. Thus, each vertex contains all the degrees of freedom of a lattice gauge theory on a lattice of arbitrary size and each link carries an arbitrary number of copies of $SU(2)$-degrees of freedom. These degrees of freedom are naturally ordered by (1) the lattice size for the vertex degrees of freedom and (2) the number of copies of $SU(2)$. This suggests the construction of the algebra of vertex-observables as the limit of the observable algebras of finite lattice gauge theories together with their canonical Hilbert space representation and similarly for the link degrees of freedom, which allows us to solve consistency condition \ref{consistency} in the obvious way of embedding a sequence of sublattices into a sequence of lattices. There is evidence that the pull-back of lattice observables under this construction is a quantum embedding in the sense of \cite{tim-reduction}, furnishing the morphisms in the then defined category {\bf LGT} of the lattice gauge theories with extra degrees of freedom.

Now we have to specify the ''irrelevant degrees of freedom''. Given a physical Hamiltonian $H_{phys.}$, we have to include all degrees of freedom that have ''observable effects'', when the initial state is given by a state that does not contain one of these degrees of freedom. This means a degree of freedom is irrelevant, if  there are no lattice measurements available that can effectively distinguish between the effect of the dynamical occurrence of a particular degree of freedom and its time evolution and an effective lattice state and it time evolution of the effective lattice state. It is thus important to know the dynamics in order to determine the ''relevance'' of a degree of freedom, so we have to postpone the discussion of ''relevance'' until after the definition of a suitable dynamics.\footnote{The usual effect is that ''heavy'' and ''weakly coupled'' extra degrees of freedom are ''irrelevant'', where heavy is understood w.r.t. the lattice spacing $l_o$, when both the mass and the spacing are measured in natural units.}

We do not yet claim that this construction, which we denote by $F_{G_o}:${\bf Tri}$\rightarrow${\bf LGT}, yields a functor, due to some seemingly natural yet unproven assumptions that we had to make. We are however confident that at most ''technical details'' have to be adjusted and the general picture will turn out unchanged.

Let us now generalize the construction $F_{G_o}$ to the vacuum state $\Omega_{E_o}$ of a GNS-summand, more precisely: given a cubical decomposition $\mathcal D(C)$, we want a state $\Omega^{eff}_{E_o}$ on the lattice gauge theory $LGT(\Gamma(\mathcal D))$, such that the deviations between the expectation values are small:
$$
  \langle \Omega^{eff}_{E_o}, LattObs \Omega^{eff}_{E_o} \rangle = \omega_{E_o}(I(LattObs))+ small\,corr.,
$$
here $I$ denotes the canonical embedding of a lattice observable into an element of the algebra of quantum geometry, given by mapping link observables on the lattice into the respective holomomy observable along the embedding of this link in the algebra of quantum geometry and a conjugate momentum into the flux observable through the embedding of that face of $\mathcal D$ that is dual to the momentum. A particular choice for such observables is given by a momentum squeezing $|\Psi_{E_o}(\Gamma)\rangle:=\rangle\lim_{t\rightarrow large}|\psi^\Gamma_t(0,E_o)\rangle$ of Thiemanns coherent sates\cite{gcs}, which are proven to yield the correct expectation values \cite{Thiemann-winkler}. We will assume, although we cannot prove it due to the lack of a physical Hamiltonian, that the extra vertex and link degrees of freedom are dynamically irrelevant, which should hold at least for classical static solutions $E_o$ of Einstein's equations. Under these condition it turns out that the construction $F_{E_o}$ yields a family of lattice states:
\begin{equation}
  \{|\Psi_{E_o}(\Gamma)\rangle:\Gamma(\mathcal D)\}_{\mathcal D \in comp.cub.dec.(\Sigma)},
\end{equation}
one for each cubical decomposition of a compact subset of $\Sigma$. Given any surface $S$, the family has the feature that the expectation values
$$
  \langle \Psi_{E_o}(\Gamma), A(S) \Psi_{E_o}(\Gamma)\rangle = A_{E_o}(S)+small\,corr.,
$$
whenever $\Gamma=\Gamma(\mathcal D)$ and $S$ can be decomposed into faces in $\mathcal D$. The analogous statement holds for the expectation values of volumes of regions. To make a connection with the F/LOST representation of the algebra of quantum geometry, let us consider the lattice states as states on the embedded lattice in the F/LOST-representation and put this observation on its head: Consider a (projective) family of states $\{\psi_{E_o,\mathcal S,\mathcal R}\}_{\mathcal S,\mathcal R}$ indexed by finite sets of surfaces $\mathcal S$ and regions $\mathcal R$ in $\Sigma$ in the F/LOST-representation, such that for any countable set of surfaces and regions the expectation values
$$ \langle \psi_{E_o,\mathcal S,\mathcal R}, A(S) \psi_{E_o,\mathcal S,\mathcal R}\rangle=A_{E_o}(S) \, and \, \langle \psi_{E_o,\mathcal S,\mathcal R}, V(R) \psi_{E_o,\mathcal S,\mathcal R}\rangle=V_{E_o}(R), $$
for all $S \in \mathcal S$ and all $R \in \mathcal R$. This family is partially ordered (using the joint subset relation) and projective, so one can heuristically consider the projective limit:
\begin{equation}
  \Psi_{E_o}:=\lim_{\leftarrow (\mathcal S,\mathcal R)} \psi_{E_o,\mathcal S,\mathcal R},
\end{equation}
which does not exist in the F/LOST-representation. We can however define a state through the vacuum expectation values, which will then coincide with $\omega_{E_o}$. 

This situation is reminiscent of a Bose condensate ground state of a free scalar field theory: Given a particular ground state density $\rho_o$, one can consider the thermodynamic limit $\Lambda \rightarrow \infty$ of a family of grand canonical states
$$
  \omega_{\Lambda,\beta,\mu}(a)|_{\rho(\beta,\mu)=\rho_o}:=\frac{Tr(e^{-\beta H+\mu N} a)}{Tr(e^{-\beta H+\mu N})}V,
$$
where the inverse temperature $\beta$ and the chemical potential $\mu$ are adjusted so they yield the expectation value $\rho_o$ in an increasing region as $\Lambda \rightarrow \infty$. Generally there is no element in the Hilbert space of the free theory that reproduces this limit and one has to preform the GNS-construction w.r.t. the state defined through
\begin{equation}\label{bec}
  \omega_{\rho_o}(a):=\lim_{\Lambda \rightarrow \infty}\omega_{\Lambda,\beta,\mu}(a)|_{\rho(\beta,\mu)=\rho_o}.
\end{equation}
The r\^{o}le of $\rho_o$ is similar to the r\^{o}le of $E_o$ in the new representation of the algebra of quantum geometry, so one can view the state $\omega_{E_o}$ as a ''local condensate of quantum geometry''. The term ''local'' is motivated by: Given any open set $O \subset \Sigma$ (that is contained in a compact set) of arbitrarily small size one can preform the limit construction $  \Psi_{E_o}:=\lim_{\leftarrow (\mathcal S,\mathcal R)} \psi_{E_o,\mathcal S,\mathcal R}$ for surfaces and regions in $O$ and one obtains that the projective limit does not exist in the F/LOST-representation, but there exists a state $\omega_{E_o|_O}$, which can be defined similar to equation \ref{bec}. 

One can also draw a similarity between the GNS-Hilbert space constructed from the BEC ground state and the $E_o$-GNS-Hilbert space: The GNS-construction in the BEC-case yields a Hilbert space that contains states that have ground state density $\rho_o$ everywhere, except for quantum fluctuations around the condensate that vanish at infinity. These fluctuations can be characterized in precisely the way that the effect of a state was characterized and the ground state density $\rho_o$ has the r\^{o}le of the essential geometry. So if we adopt the interpretation of $\omega_{E_o}$ as a state describing a local condensate of geometry, we are let to view the spin network functions as quantum fluctuations around the geometric condensate. The interpretation of the F/LOST-ground state $\omega_o$ is in light of these considerations rather simple as the $\omega_{E_o}$ state with totally degenerate $E_o=0$.

Let us now return to the construction $F_{E_o}$, which we want to generalize to an arbitrary state in the $E_o$-GNS-Hilbert space, so we can find a family of lattice gauge theory-states that describe the dynamically relevant degrees of freedom of the $E_o$-state. Let us again assume that the geometric background corresponds to a classical static solution $E_o$ of Einstein's equations and that this is sufficient for not producing extra lattice degrees of freedom. Let us consider a sate $\pi(Cyl_\alpha) \Omega_{E_o}$ and a cubical decomposition $\mathcal D(C)$ of a compact subset $C \subset \Sigma$. It is our aim to construct a state on the lattice gauge theory on $\Gamma(\mathcal D(C))$, such that there are no lattice measurements that deviate significantly from the corresponding embedded measurements in the GNS-representation. Since there are no restrictions on the cylindrical function $Cyl$ or on its graph $\alpha$, we cannot rule the dynamical relevance of any part of it out, so we have to construct a state that contains all degrees of freedom. A particular construction is:
\\
(1) Construct a state $\Psi_{E_o}(\Gamma)$, that captures the essential geometry of $\pi(Cyl_\alpha)\Omega_{E_o}$ precisely as previously done for $\Omega_{E_o}$. Construct the effective state $Cyl^\prime$ through the multiplication operators:
$$ Cyl^\prime := \Psi_{E_o}(\Gamma) Cyl_\alpha $$
This procedure is supposed to absorb the $E_o$-geometry inside the cell and let is reappear on scales larger than the cell.\\
(2) Denote the restriction of $Cyl^\prime$ to a cell $c \in \mathcal D(C)$ by $Cyl_\alpha|_c$. Use the smallest cubical lattice that supports a graph $\gamma_c$ that is topologically equivalent to $\alpha|_c$. Use $\phi_c:\gamma_c\mapsto \alpha|_c$ and map the state $\phi_c^* Cyl^\prime|_c$ into the vertex Hilbert space precisely as prescribed in the construction $F_{E_o}$ for lattice gauge theories on refined lattices. If there is a set $\Gamma_c(\alpha)$ of minimal lattices, then preform this construction for all minimal lattices $\gamma_c^i\in \Gamma_c(\alpha)$ and construct
$$ \frac{1}{|\Gamma_c(\alpha)|} \sum_{\gamma_c^i \in \Gamma_c(\alpha)} \phi_{c,i}^* Cyl^\prime|_c $$ 
in the vertex Hilbert space. \\
(3) The graph $\alpha$ will in general penetrate the face $f \in \mathcal D(C)$ $n$ times and the $n$ spin quantum numbers of the penetration are mapped into the link Hilbert space.\\
(4) This procedure is only well defined, if all vertices of $\alpha$ are in the interior of a cell and if all edges penetrate the faces transversally. To resolve the ''degenerate'' cases, we have to define for each node in $\mathcal D$ to belong to the inside of a link, for each link to belong to the inside of a face and for each face to belong to the inside of a cell. Loosely speaking this procedure ''pinches $\alpha$ a bit, so the degenerate topological relation is deformed into a general topological relation.''

Let us briefly notice the extra effective vertex degrees of freedom that arise from this construction:\\
(1) Couplings to the background (which are due to the enlargement of the solution space of the Gauss-constraint in the $E_o$-representation and not an effect of the construction $F_{E_o}$)\\
(2) The topology-class of $\alpha|_c$ and the cylindrical function of this topology class\\
(3) High valent vertices (since the dual graph contains vertices of at most valence six)\\
(4) Knotting with the effective lattice state graph

The extra effective link degrees of freedom arising from this construction are:\\
(1) multiple penetration of a face by possibly different edges\\
(2) linking information of the penetrations and $\alpha|_c$ for adjacent cells $c$
\begin{figure}
  \center
  \includegraphics[width=4cm]{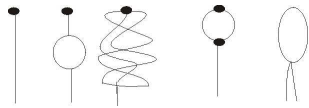}\\
  \caption{Some of the ''extra vertex degrees of freedom'' arising in our construction. Notice that these ''extra degrees of freedom'' do not occur in a lattice version of Ashtekar gravity. The relevance of the degrees of freedom is decided by the dynamics. The usual argument from perturbatively renormizable QFT that the new degrees of freedom on a finer lattice effectively decouple from degrees of freedom on the coarser lattice is not obvious in our case due to the occurrence of couplings to the background, that may occur between extra degrees of freedom on the finer lattice in the $E_o$-representation.}
\end{figure}

The construction again yields a family of lattice states
$$
  \{|\Psi_{E_o}(Cyl,\Gamma)\rangle:\Gamma(\mathcal D)\}_{\mathcal D \in comp.cub.dec.(\Sigma)},
$$
that describe the state $\pi(Cyl)\Omega_{E_o}$ up to small corrections on the respective lattices $\Gamma(\mathcal D)$. With the interpretation of the states $\pi(Cyl)\Omega_{E_o}$ as a condensate of quantum geometry, we can view the family $|\Psi_{E_o}(Cyl,\Gamma)\rangle$ as states in the F/LOST-representation in which the effect of the geometric condensate is integrated out and absorbed into the state on scales larger than the lattice spacing $l_o$, that we assumed to be ''almost regular'' from the onset. We can thus view our construction as a scale dependent (i.e. $l_o$ dependent) map that maps the essential geometry $E_o$ into an effective state in the ''fundamental'' F/LOST-representation. We can also try to put this on its head and assert that a particular representative $|\Psi_{E_o}(Cyl,\Gamma)\rangle$ in the F/LOST-representation is the true quantum state of our geometry and that the state $\pi(Cyl) \Omega_{E_o}$ is only an effective state that describes the ''smooth part'' of the quantum geometry of $|\Psi_{E_o}(Cyl,\Gamma)\rangle$ up to a scale $l_o$ given by the average lattice spacing of $\Gamma$. The problem with this interpretation is however, that there is no sense of ''nearness'' in the F/LOST-representation.

Let us now briefly discuss the possible dynamics for these states: Ideally one would have a master constraint for the F/LOST-representation and induce one for the $E_o$-representation through requiring that the following diagram commutes:
$$
      \begin{array}{crccclc}
        &|(E_o)\rangle_f&&&&F_{E_o}(\mathcal D_f)&\\
        F/LOST+\hat{M}&\rightarrow&LGT_{f}&+&\hat{M}_{f}&\leftarrow&E_o-rep.+dynamics\\
        &\searrow&&\downarrow&&\swarrow&\\
        &|(E_o)\rangle_c&LGT_{c}&+&\hat{M}_{c}&F_{E_o}(\mathcal D_c),&
      \end{array}
$$
where the diagram reads as follows: We start in the top line and add to the F/LOST-representation a lattice state $|(E_o)\rangle_f$ on a graph $\Gamma(\mathcal D_f)$, that is dual to a fine cubical decomposition $\mathcal D_f$. Applying an adaption of the construction $F_{E_o}$ with $E_o=0$ to this leads to a lattice gauge theory on a fine lattice. We tried to construct $F_{E_o}$, such that the application of $F_{E_o}$ to the $E_o$-state on the left (using the same fine cubical decomposition $\mathcal D_f$) yields the same lattice gauge theory (with extra degrees of freedom). But a master constraint on the left yields a constraint surface, that induces a constraint surface on the right.

The lower line describes the analogous construction for a coarsening $\mathcal D_c$ of $\mathcal D_f$. The nontrivial consistency condition is that the construction $F$ between the lattice gauge theories yields the lattice gauge theory with extra degrees of freedom on the coarser lattice. Since the dynamics of the F/LOST-representation is still disputed, we may take a different route and forget about the F/LOST-side of the diagram and try to ''invent'' a consistent family of master constraints for the lattice gauge theories with extra degrees of freedom in the middle.

This letter is necessarily incomplete, particularly the proofs of our statements where not carried out in detail. The missing details about the volume operator will follow in \cite{tim-volume}, details about the algebra, the non-vacuum state and the GNS-representation will follow in \cite{tim-algebra}, details about the construction of effective field theories will follow in \cite{tim-effective} and the precise mathematical formulation of the category {\bf LAT} and the functor $F_{E_o}$ as well as the possibility of the definition of a consistent dynamics for these effective field theories are currently under investigation.

\subsubsection*{Acknowledgements:} One third of this work was supported by the Deutsche Forschungsgemeinschaft. Another third of this work profited from an invited visit to the Perimeter Institute. I am grateful for helpful discussions with Bianca Dittrich in particular and useful discussions with Lee Smolin, Laurent Freidel, Thorsten Ohl, Martin Bojowald and Jonathan Engle. I am also thankful for my education at the Benedictine Abbey of M\"unsterschwarzach, which amongst many other useful things taught me austerity, which was necessary to complete this work.

\end{document}